\documentclass[twocolumn,english,prl,superscriptaddress]{revtex4}
\usepackage[T1]{fontenc}
\usepackage[latin9]{inputenc}
\usepackage{graphicx}
\usepackage{nomencl}

\providecommand{\makenomenclature}{\makeglossary}
\makenomenclature



\usepackage{babel}

\begin{document}

\title{Dark Matter: The Leptonic Connection}

\author{Qing-Hong Cao}

\affiliation{HEP Divison, Argonne National Laboratory, Argonne, IL 60439}
\affiliation{Enrico Fermi Institute, University of Chicago, Chicago, 
IL 60439}

\author{Ernest Ma}

\affiliation{Department of Physics and Astronomy, University of California, 
Riverside, CA 92531}

\author{Gabe Shaughnessy}

\affiliation{HEP Divison, Argonne National Laboratory, Argonne, IL 60439}
\affiliation{Department of Physics and Astronomy, Northwestern 
University, Evanston, IL 60208}

\begin{abstract}
Recent observatons of high-energy positrons and electrons by the PAMELA and 
ATIC experiments may be an indication of the annihilation of dark matter into 
leptons and not quarks.  This leptonic connection was foreseen already some 
years ago in two different models of radiative neutrino mass.  We discuss 
here the generic interactions $(\nu \eta^0 - l \eta^+) \chi$ and $l^c \zeta^-
\chi^c$ which allow this to happen, where $\chi$ and/or $\chi^c$ are 
fermionic dark-matter candidates. We point out in particular the importance 
of $\chi \chi \to l^+ l^- \gamma$ to both positron and gamma-ray signals within this framework.
\end{abstract}

\maketitle

\underline{\it Introduction}~:~ Dark matter (DM) is widely 
recognized as a necessary component of the Universe, but its nature remains 
unknown.  A possible hint to solving this mystery is the recent observation 
of high-energy positrons and electrons by the PAMELA~\cite{pamela1,pamela2} 
and ATIC~\cite{atic} experiments without any accompanying evidence of 
antiprotons.  Consider the interactions
\begin{equation}
f (\nu \eta^0 - l \eta^+) \chi + f' l^c \zeta^- \chi^c + h.c.
\end{equation}
in addition to those of the standard model (SM) of quarks and leptons, 
where the new scalars $\eta^0, \eta^+, \zeta^-$, and the new fermions 
$\chi, \chi^c$ are odd under an exactly conserved $Z_2$ symmetry, while 
all SM particles are even.  Assume also the conservation of lepton number 
$L$ so that $\chi$ (or $\eta$) has $L=-1$, and $\chi^c$ (or $\zeta$) has 
$L=1$.  To accommodate nonzero neutrino masses, the usual seesaw mechanism 
may be invoked with the term $(\nu \phi^0 - l \phi^+) N^c$, where $\Phi = 
(\phi^+,\phi^0)$ is the SM Higgs doublet and $N^c$ is a neutral singlet 
fermion with $L=-1$, both of which are even under $Z_2$.  A large Majorana 
mass for $N^c$ will then break $L$ to $(-)^L$ and allow $\nu$ to acquire a 
naturally small Majorana mass, as is well-known.  Similarly, a Majorana mass 
for $\chi$ may also break $L$ to $(-)^L$, in which case the quartic scalar 
term $(\lambda_5/2) (\Phi^\dagger \eta)^2 + h.c.$ is allowed and a neutrino 
mass is generated in one-loop order~\cite{m98}.  If $N^c$ is absent, $m_\nu$ 
will be generated solely by particles which are odd under $Z_2$, as proposed 
already three years ago~\cite{m06-1}, and may be called {\it scotogenic}, 
i.e. caused by darkness. As for the $f'$ interaction of Eq.~(1), it was 
used in a three-loop model of neutrino mass~\cite{knt03} and in two models 
of leptogenesis~\cite{fhm06,hkmr07}.

Whereas the interactions of Eq.~(1) are motivated by neutrino mass and 
leptogenesis, we adopt the viewpoint in this paper that the couplings $f$ 
and $f'$, as well as the masses of the new particles, are unconstrained 
parameters, to be explored for their DM properties.  For definiteness, we 
will study the case where $f'=0$ and $\chi$ is Majorana or Dirac.  Our 
results are easily adaptable to the case where $f=0$ and $\chi^c$ is 
Majorana or Dirac, and to the case where both $f$ and $f'$ are nonzero, 
with $\chi$ and $\chi^c$ forming a Dirac fermion.

\underline{\it Elaboration}~:~ The $f$ interaction of Eq.~(1) has 
been studied before.  The case of $L=-1$ for $(\eta^+,\eta^0)$ 
and $L=0$ for $\chi$ is the leptonic Higgs model~\cite{m01}.  There $L$ is 
broken explicitly by the soft term $\Phi^\dagger \eta$, so that a small 
$\langle \eta^0 \rangle$ is obtained which allows $\nu$ to pair up with 
$\chi (= N^c)$ to acquire a Dirac mass. Together with the allowed Majorana 
mass for $\chi$, a seesaw mass for $\nu$ is obtained.  The case of $L=0$ 
for $(\eta^+,\eta^0)$ and $L=-1$ for $\chi$ where both are odd under an 
exactly conserved $Z_2$ is the prototype model of scotogenic neutrino mass 
~\cite{m06-1}.  In this case, the quartic scalar term $(\lambda_5/2)
(\Phi^\dagger \eta)^2 + h.c.$ is allowed, which splits $\eta^0$ into two 
particles, $Re(\eta^0)$ and $Im(\eta^0)$, with different masses~\cite{dm78}.  
This allows the lighter of the two to be considered as dark matter 
\cite{m06-1,bhr06,lnot07,glbe07}. The collider signature of this scenario 
has also been discussed~\cite{cmr07}, as well as its cosmological 
implications~\cite{Barger:2008jx}.

If $Re(\eta^0)$ or $Im(\eta^0)$ is dark matter, then its annihilation to 
a pair of gauge bosons or the SM Higgs boson will result in both quarks and 
leptons.  If $\chi$ or $\chi^c$ is dark matter, then only leptons are 
expected.  If $f'=0$, then $\chi$ may annihilate to neutrinos and charged 
leptons through the exchange of $\eta^0$ and $\eta^+$.  If $f=0$, then 
$\chi^c$ may annihilate only to charged leptons through the exchange of 
$\zeta^-$.  In these two cases, we can assume $\chi$ or $\chi^c$ to be 
either Majorana or Dirac.  In the case $f \neq 0$ and $f' \neq 0$, the 
natural scenario is that $\chi$ and $\chi^c$ together form a Dirac 
fermion.  To avoid flavor-changing leptonic interactions, such as $\mu \to 
e \gamma$, which is an intrinsic problem \cite{kms06} in models of 
scotogenic neutrino mass if $\chi$ is considered as dark matter, we will 
assume for simplicity that $\chi$ couples only to $e$ and $\nu_e$.

\begin{figure}
\includegraphics[clip,scale=0.42]{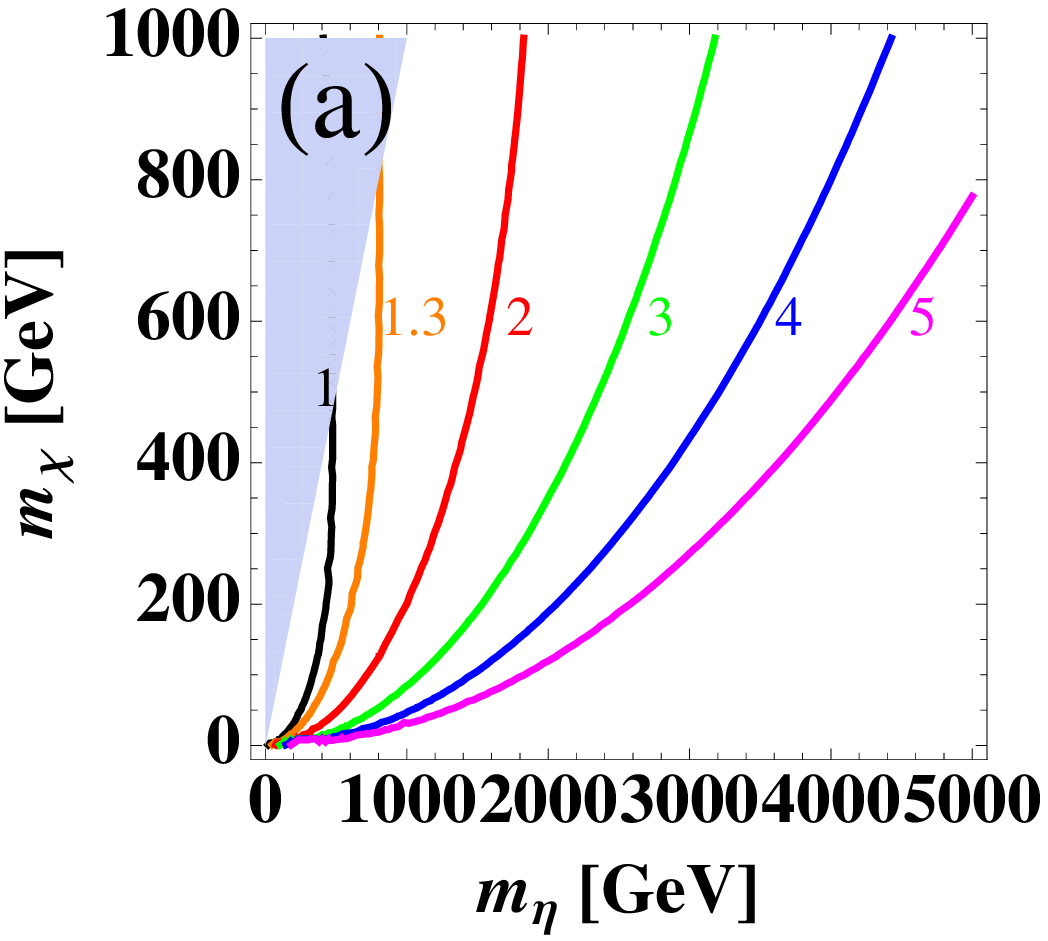}\includegraphics[clip,scale=0.4]{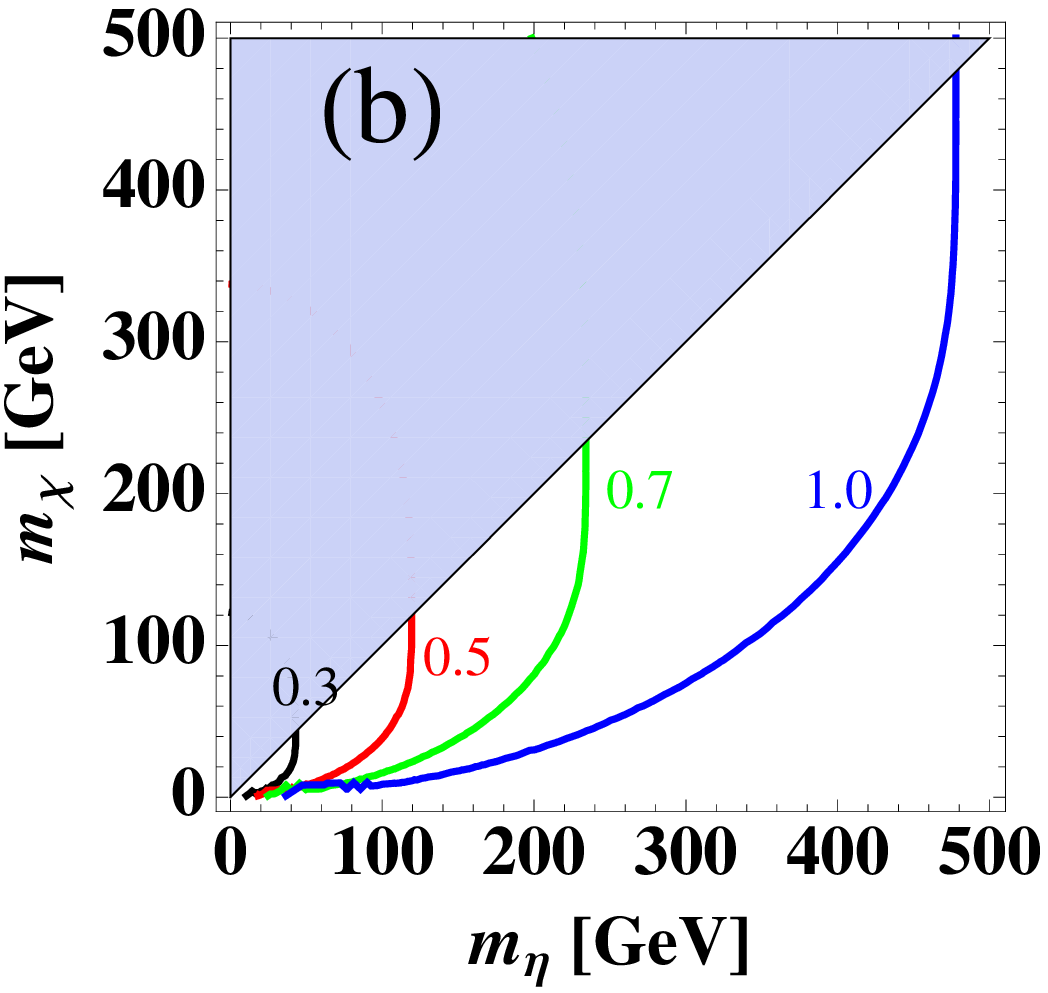}

\includegraphics[clip,scale=0.42]{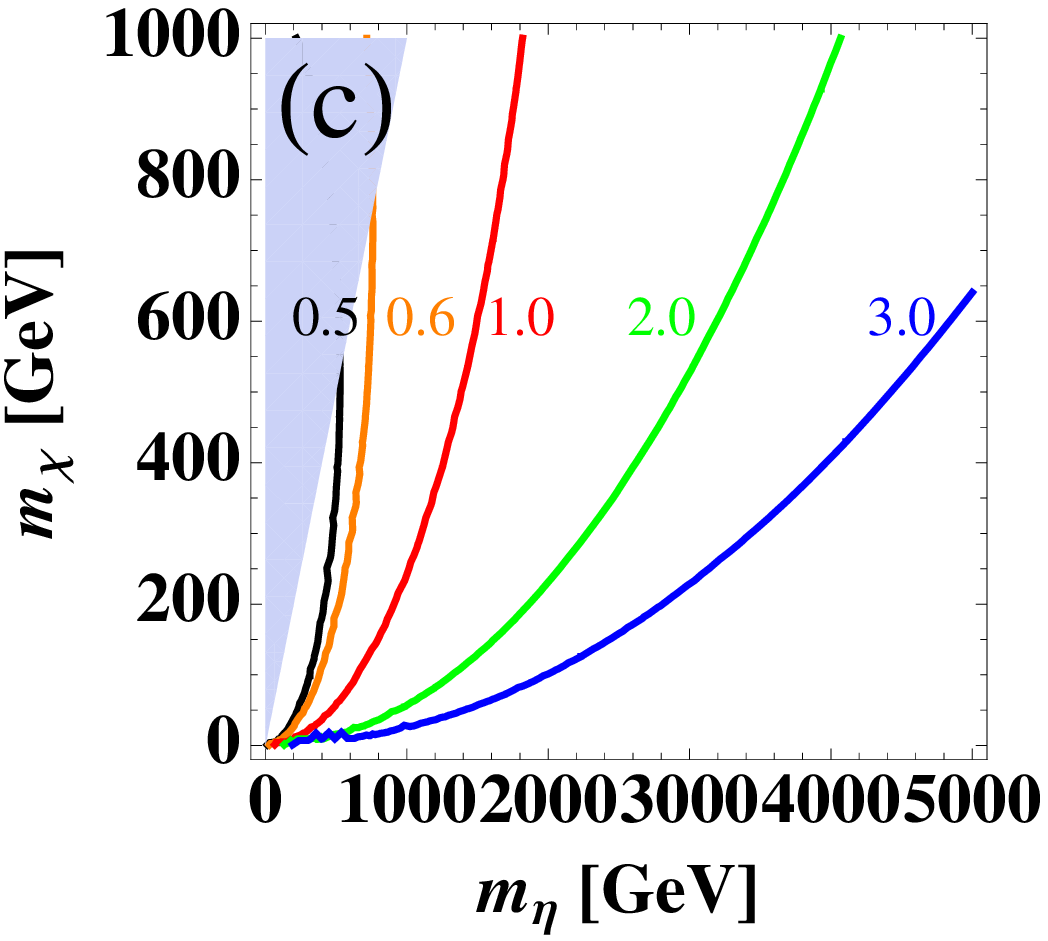}\includegraphics[clip,scale=0.42]{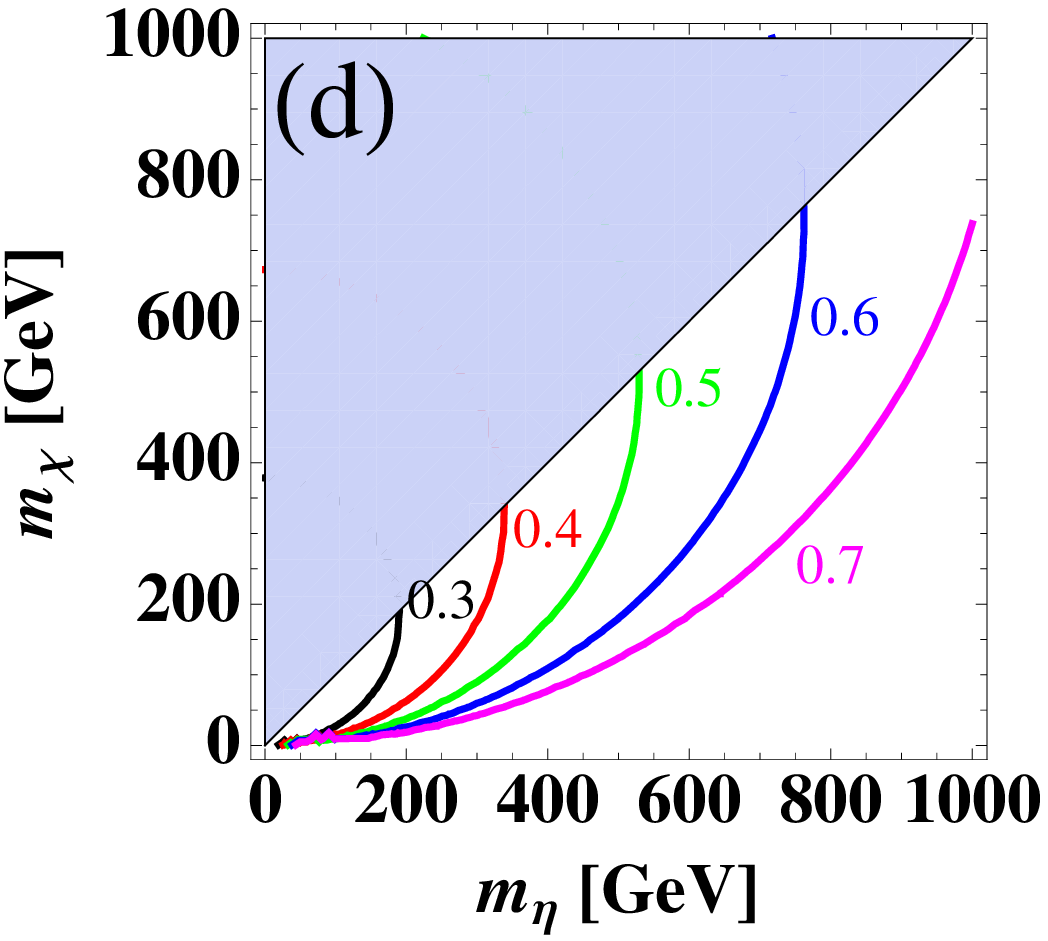}

\caption{Correlation between $m_{\chi}$ and $m_{\eta}$, giving rise to the 
correct amount of relic abundance for different Yukawa coupling strengths: 
(a) and (b) for Majorana $\chi$, (c) and (d) for Dirac $\chi$. The
shadow region where $m_{\chi}>m_{\eta}$ is excluded.\label{fig:relic}}

\end{figure}

\underline{\it Relic abundance}~:~ As mentioned earlier, we will 
consider for definiteness, only the $f$ interaction of Eq.~(1).  However, 
we will make the important distinction between whether $\chi$ is Majorana 
or Dirac.  The observed relic abundance of $\chi$ is determined by its 
thermally averaged annihilation cross section into charged leptons and 
neutrinos multiplied by its velocity at the time (or equivalently the 
temperature) of its decoupling from the SM particles in the early Universe. 
To this end, we use the well-known nonrelativistic approximation 
$\langle \sigma v \rangle = a + b v^2$.
If $\chi$ is Majorana, we have
\begin{equation}
\label{eq:majann}
{\rm (A)}~~a=0,\qquad b=\frac{f^4 r^{2}\left(1-2r+2r^{2}\right)}
{24\pi m_{\chi}^{2}},
\end{equation}
where we have assumed for simplicity equal masses for $\eta^\pm$ and $\eta^0$, 
with $r\equiv m_{\chi}^{2}/(m_{\eta}^{2}+m_{\chi}^{2})$. If $\chi$ 
is Dirac, we have
\begin{equation}
\label{eq:dirann}
{\rm (B)}~~a=\frac{f^{4}r^{2}}{16\pi m_{\chi}^{2}},\qquad b=\frac{f^{4}r^{2}
\left(11-40r+24r^{2}\right)}{384\pi m_{\chi}^{2}}.
\end{equation}

As shown in Fig.~\ref{fig:relic}, there is a strong correlation between 
$m_{\chi}$ and $m_{\eta}$ for a given $f$, in order that the correct 
amount of DM relic abundance~\cite{wmap07} be produced.  Smaller values of 
$f$ are allowed if $\chi$ is Dirac (B) rather than Majorana (A) because of 
$a=0$ in Eq.~(2).  It should also be mentioned that the coannihilation of 
$\chi$ and $\eta$ can reduce $f$ in both cases, if $m_\eta$ is only 
slightly greater than $m_\chi$~\cite{cs09}.  Here we do not consider 
this scenario.

\begin{figure}
\includegraphics[clip,scale=0.4]{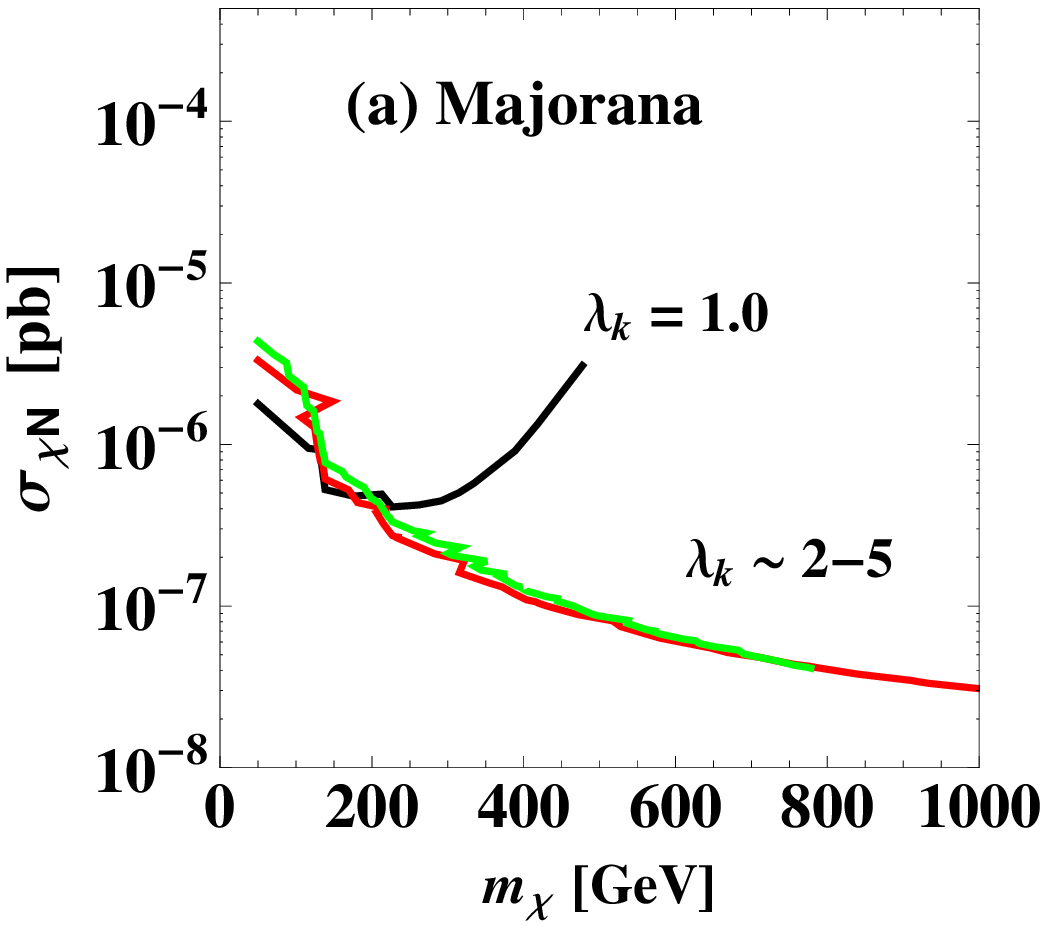}
\includegraphics[clip,scale=0.4]{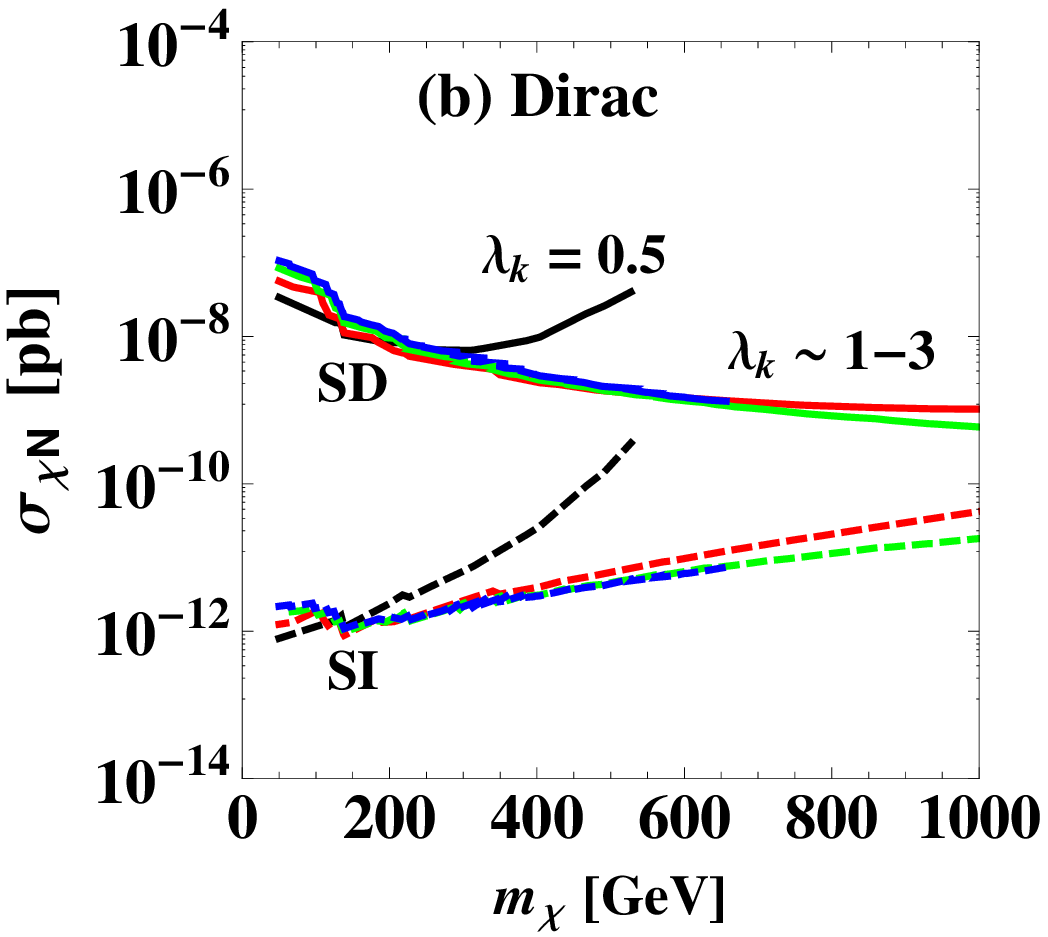}

\caption{Cross sections for the direct detection of $\chi$: (a) Majorana 
case, SD only; (b) Dirac case, both SD and SI.\label{fig:SD}}

\end{figure}

\underline{\it Direct detection}~:~ Even though $\chi$ couples 
only to leptons, it can interact with quarks through its one-loop effective 
coupling to the $Z$ boson.  It may thus be detectable in the next-generation 
direct-search experiments for dark matter using nuclear recoil.  If $\chi$ 
is Majorana, its effective interaction with quarks is given by
\begin{equation}
\mathcal{L}_{\rm A}=\frac{\mathcal{G}}{m_{Z}^{2}}(\bar{\chi}
\gamma^{\mu}\gamma_{5}\chi)(\bar{q}\gamma_{\mu}\gamma_{5}q),
\end{equation}
where $\mathcal{G}$ is the loop-induced form factor. Fig.\ \ref{fig:SD}(a)
shows the spin-dependent (SD) cross section ($\sigma_{SD}$) as a function 
of $m_{\chi}$ with $f=1.0$ (black) and $f\sim2-5$ (green, red), where 
$m_{\eta}$ is determined by the relic abundance.  It is clear that the SD 
cross section is well below the current experimental bounds, ${\cal O}(10^{-2})$ pb, owing to the 
loop suppression, but may be detectable in future experiments. If 
$\chi$ is Dirac, there exist both vector and axial-vector coupllings, thus
\begin{equation}
\mathcal{L}_{\rm B}=\frac{\mathcal{G}'}{m_{Z}^{2}}(\bar{\chi}\gamma^{\mu}
\chi)(\bar{q}\gamma_{\mu}q)+\frac{\mathcal{G}''}{m_{Z}^{2}}
(\bar{\chi}\gamma^{\mu}\gamma_{5}\chi)(\bar{q}\gamma_{\mu}\gamma_{5}q).
\end{equation}
This means that there is a spin-independent (SI) cross section which can be 
enhanced by coherent effects as well as a spin-dependent one.  
Fig.~\ref{fig:SD}(b) shows both the SD (solid curves) and SI (dashed curves) 
cross sections.  In this case, both may be outside the reach of experiments 
in the near future.

\underline{\it Observation of positrons}~:~ 
The measurement of secondary particles coming from the annihilation of dark 
matter in the halo of the galaxy provides a promising way of deciphering 
its nature.  If positrons are produced and propagate through the galaxy, their 
spectrum is distorted as they pass through the turbulent galactic magnetic 
fields.  This propagation can be described by a diffusion-energy-loss 
process~\cite{Delahaye:2007fr}.  The resulting flux seen at Earth is scaled 
by an overall normalization (boost) factor, due to the unknown level of 
clumping of dark matter at the positron source.  Recently, two experimental 
collaborations, PAMELA and ATIC, have reported an excess of high-energy 
positrons and electrons.  These confirm earlier results from HEAT and AMS-01, 
raising the exciting possibility that dark matter annihilates either directly, 
or indirectly to positrons, but not to antiprotons.
 
The PAMELA collaboration observed an excess well above the expected background 
in the positron fraction at energies $10-100$ GeV~\cite{pamela1}. Many 
explanations have recently been proposed to account for this excess including 
SM extensions~\cite{Huh:2008vj,ArkaniHamed:2008qn,Pospelov:2008jd,Hisano:2008ti,Gelmini:2008vi,Fairbairn:2008fb,Nelson:2008hj,Cholis:2008qq,Harnik:2008uu,
Feldman:2008xs,Bai:2008jt,Fox:2008kb,Baek:2008nz,Zurek:2008qg,Shepherd:2009sa,Park:2009cs}, decaying DM~\cite{Chen:2008dh,Yin:2008bs,Nardi:2008ix} and non-DM 
astrophysical sources~\cite{Hooper:2008kg,Profumo:2008ms}.  Due 
to the abrupt rise in the positron fraction with increasing energy, the 
resulting positron spectrum injected into the halo is expected to be quite 
hard, indicating either direct annihilation to $e^+e^-$, or states such as 
$\mu$ and $\tau$ leptons which give off energetic secondary positrons. This 
has been quantified in recent studies~\cite{Cirelli:2008pk,Barger:2008su,Cholis:2008hb}. 
To be consistent with PAMELA data, the DM masses favored are of the order a 
few hundred GeV, given a marginalization over all possible annihilation modes 
which result in positrons~\cite{Barger:2008su}.  

Recently, the ATIC~\cite{atic} collaboration also observes an excess of 
high-energy positrons and electrons.  In addition, the data exhibit an 
excess in the $\Phi_{e^+}+\Phi_{e^-}$ spectrum in the range 300-800 GeV.  
This result at first sight seems to be at odds with the DM candidate favored 
by the PAMELA data.  However, a combined fit shows that a 700-850 GeV DM 
candidate is consistent with both PAMELA and ATIC if only charged leptons 
are allowed.  

\begin{figure}[htb]
\includegraphics[clip,scale=0.4]{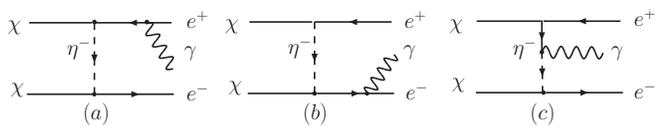}
\caption{Feynman diagrams of the bremstrahlung annihilation process.}
\label{fig:fd}
\end{figure}
\underline{\it Dark matter bremsstrahlung}~:~ 
The preference for DM annihilations to hard positrons suggests the scenario 
of Eq.~(1).  In this case, whether $\chi$ is Majorana or Dirac has a striking 
impact on the resulting positron spectrum.  If it is Majorana, the annihilation 
cross section is helicity suppressed, leaving only the p-wave component in 
Eq.~\ref{eq:majann}.  However, the velocities at which dark matter annihilates 
today (as opposed to the early Universe) are so small, $v/c\approx 10^{-3}$, 
the p-wave term is even more negligible.  Thus at tree level, annihilations 
to $\tau$ leptons are dominant because the s-wave terms are scaled by the 
mass-squares of fermions to which the DM pairs annihilate.  However, the 
helicity suppression may still be severe enough that the bremsstrahlung 
process $\chi \chi \to e^+ e^- \gamma$ can be much more important 
(see Fig.~\ref{fig:fd}).  This effect has been studied in the 
context of supersymmetry in the stau-coannihilation region for the PAMELA 
excess~\cite{Bergstrom:2008gr}.  In the extreme nonrelativistic
limit $v\to0$, the cosmic positron spectrum is given by~\cite{Bergstrom:2008gr}
\begin{eqnarray}
 && \frac{d\sigma}{dE_{e^{+}}}\Biggr|_{v\to0}=~~\frac{\alpha f^{4}}
{256\pi^{2}m_{\chi}^{2}}\frac{1}{\left(2x+\mu-1\right)^{2}}~~ \times \nonumber \\
 && \Biggl\{\left(4\left(1-x\right)^{2}-4x(1+\mu)+3(1+\mu)^{2}\right)
\log\frac{1+\mu}{1+\mu-2x} \nonumber \\
 && \,\,\,\,-\left(4(1-x)^{2}-x(1+\mu)+3(1+\mu)^{2}\right)
\frac{2x}{1+\mu}\Biggr\},
\end{eqnarray}
with $x\equiv E_{e^{+}}/m_{\chi}$ and $\mu=m_{\eta^{\pm}}/m_{\chi}$.
\begin{figure}
\includegraphics[clip,scale=0.4]{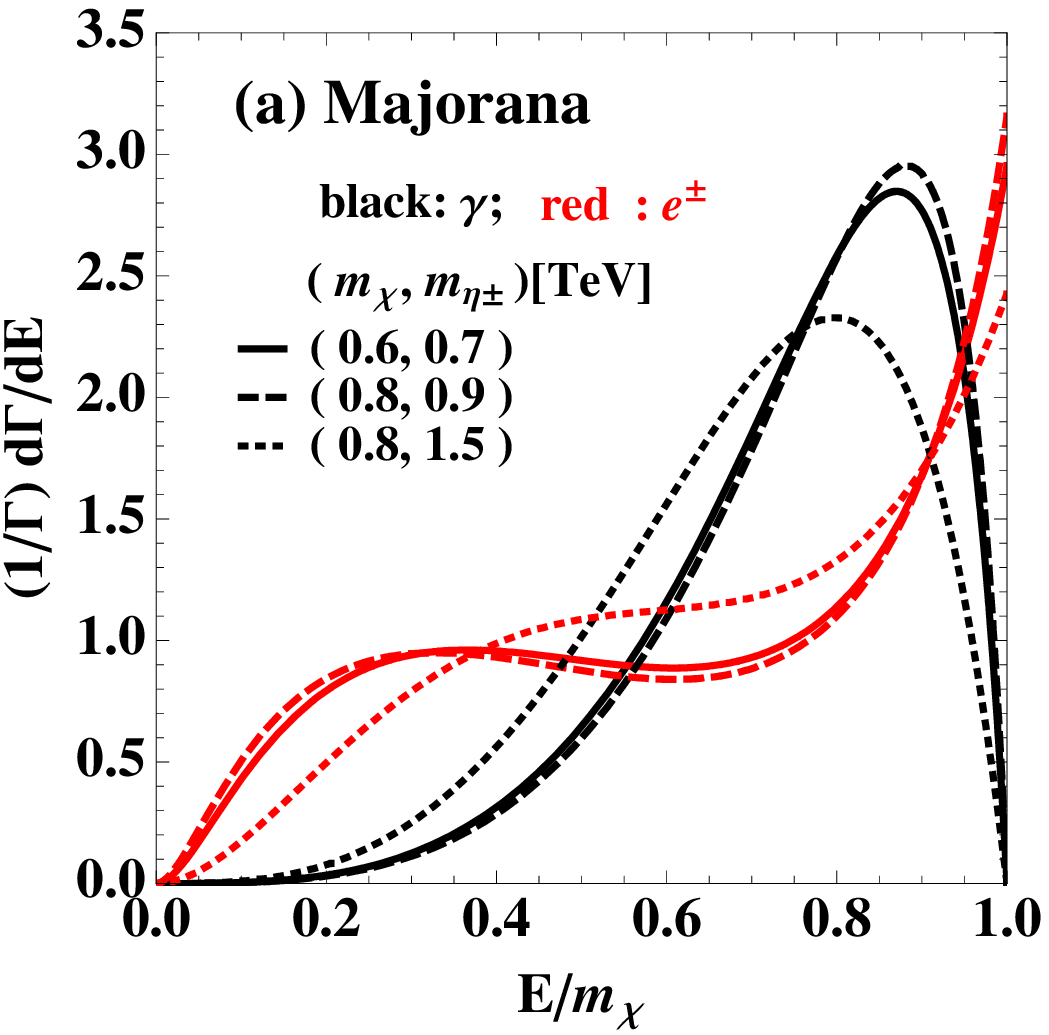}\includegraphics[clip,scale=0.4]{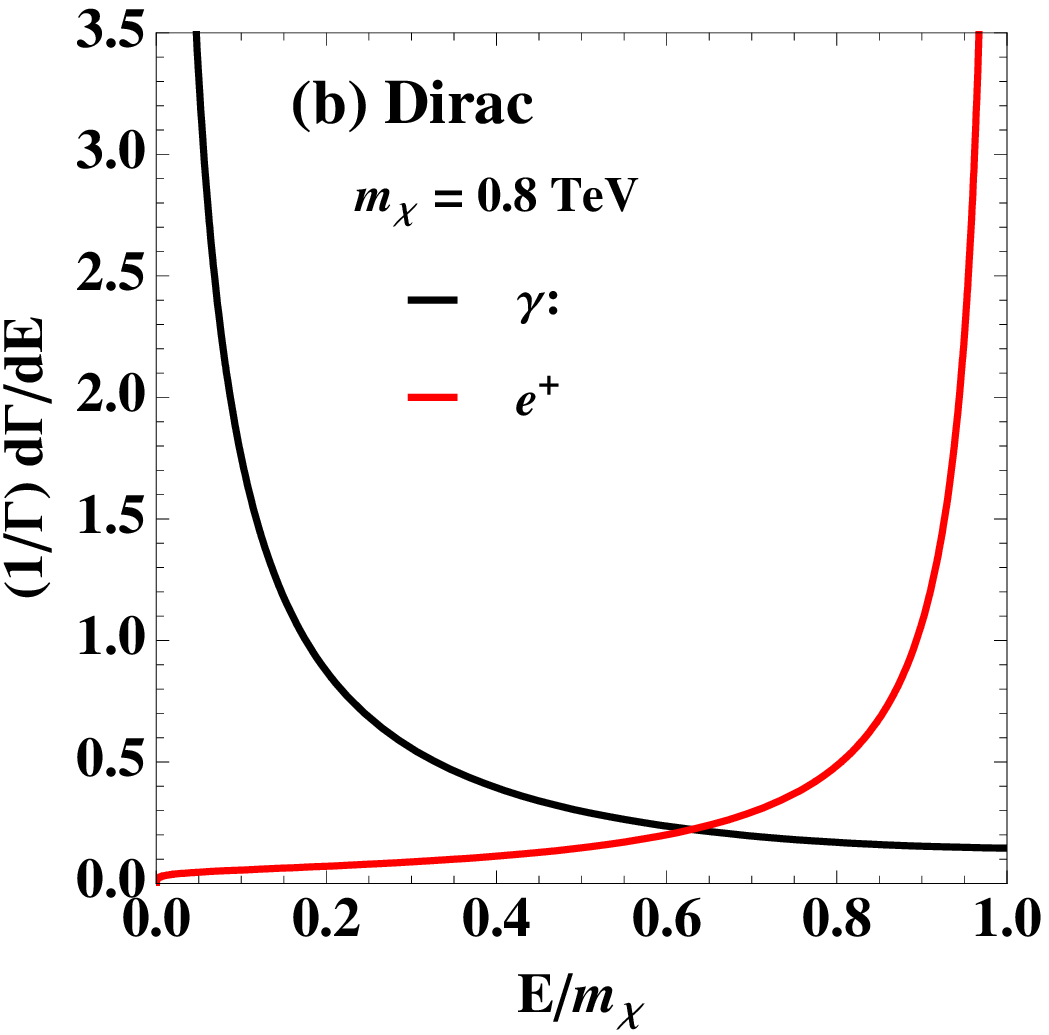}
\caption{Energy spectrum of the gamma-ray (black curve) and positron (red curve)
in the limit of $v\to0$ for $\chi \chi \to e^+ e^- \gamma$ with various values of $m_{\chi}$ and 
$m_{\eta}$.\label{fig:energy_spectrum}}
\end{figure}
The resulting positron spectrum can exhibit a sharp peak near the kinematic 
endpoint (red curves) in Fig. \ref{fig:energy_spectrum}.  Depending on 
the relative masses of $\chi$ and $\eta$, interference effects can 
distort the positron spectrum, yielding a secondary peak at lower energies. 

The associated photon in this process may also be observable with the FERMI 
gamma-ray telescope~\cite{Morselli:2008zz}.  With an upper energy range of 
$\sim$300 GeV, FERMI should be well positioned to catch a glimpse of the 
photon signal.  The energy spectrum of the gamma-ray is given by
\begin{eqnarray}
 && \frac{d\sigma}{dE_{\gamma}}\Biggr|_{v\to0}=~~\frac{\alpha f^{4}}
{256\pi^{2}m_{\chi}^{2}}\frac{y-1}{\left(1+\mu-y\right)^{2}}~~ \times \nonumber \\
 && \Biggl\{\frac{2\left(1+\mu\right)\left(1+\mu-2y\right)}
{1+\mu-y}\log\frac{1+\mu-2y}{1+\mu} \nonumber \\
 && -\frac{4x\left[y^{2}+\left(1+\mu-y\right)^{2}\right]}
{\left(1+\mu\right)\left(1+\mu-2y\right)}\Biggr\},
\end{eqnarray}
with $y\equiv E_{\gamma}/m_{\chi}$. The resulting gamma-ray spectrum from the 
bremsstrahlung photon (black curves) in the Majorana case peaks before the 
endpoint, then abruptly terminates.  With increasing $m_{\eta^{\pm}}$ the peak 
positions are shifted to the low-energy regime, see the dashed curves, i.e.
the positron and photon become softer.  

If $\chi$ is instead Dirac, the annihilation cross section is dominated 
by the s-wave component in Eq. \ref{eq:dirann}.  In this case, the DM pairs 
tend to annihilate to fermions more democratically.  Further, since there is 
no suppression, the impact of the bremstrahlung process on the positron 
energy spectrum is negligible.  Due to the strong suppression 
of the annihilation cross section to fermion pairs in the Majorana case, the 
boost factors required to give the same spectra seen by PAMELA are typically 
several orders of magnitude larger than in the Dirac case.  On the other hand, 
the photon spectrum from bremstrahlung is much softer  than 
in the Majorana case.  Since the charged leptons in the final state are
relativistic, the radiated photon predominately moves collinearly with
the charged leptons, i.e. the \textquotedblleft{}fi{}nal
state radiation\textquotedblright{} (FSR) regime. In this kinematic limit,
the cross section factorizes into the short-distance part, 
$\sigma\left(\chi\chi^{C}\to\ell^{+}\ell^{-}\right)$,
and a universal collinear factor: 
\begin{eqnarray}
 && \frac{d\sigma(\chi\chi^{C}\to\ell^{+}\ell^{-}\gamma)}{dy} \approx 
\sigma(\chi\chi^{C}\to\ell^{+}\ell^{-}) \times \nonumber \\ 
&& \frac{\alpha e^{2}}{\pi}\frac{1+(1-y)^{2}}{y} 
\log\frac{4m_{\chi}^{2}(1-y)}{m_{\ell}^{2}}.
\end{eqnarray}
The photon energy spectrum then peaks around zero.

\begin{figure}[htb]
\includegraphics[clip,scale=0.3]{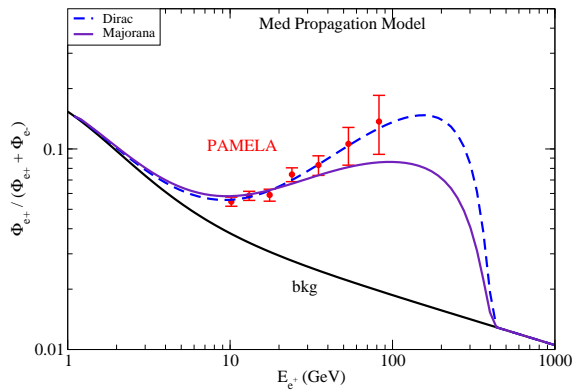}
\caption{Positron fraction in the leptonic dark matter scenario with 
$M_\chi=450$ GeV and $m_\eta=500$ GeV for $\chi$ Majorana and Dirac.}
\label{fig:pamelaillust}
\end{figure}

We illustrate, in Fig. \ref{fig:pamelaillust}, the positron fraction as seen 
at Earth after propagation effects are included in the ``Med'' propagation 
scheme following Ref.~\cite{Barger:2008su}.  Here, we take $M_\chi=450$ GeV, 
$m_\eta=500$ GeV and require the coupling $f$ such that the relic density 
is saturated.  In the Dirac case, the fit is quite good, with a boost factor 
of 91.  However, in the Majorana case, the fit is marginal as the spectrum is 
suppressed at higher energies.  Due to the helicity suppression,  a huge boost 
factor of ${\cal O}(10^6)$ is also required to fit the data in this case.
With enough precision, by comparing the correlated signals seen in high-energy 
positrons (PAMELA) and gamma-rays (FERMI), the specifics of dark matter and the 
associated exchanged particle in this scenario may be explored.

As this work is being completed, a similar paper~\cite{Bi:2009md} has  
appeared.  However, our results do not agree in the case of Majorana dark 
matter.  Specifically, it is claimed there that Majorana dark matter has a 
nonzero s-wave contribution in its annihilaton. 

\underline{\it Conclusion}~:~ The PAMELA and ATIC observations may be 
indicative of a leptonic connection in dark-matter interactions, as given 
by Eq.~(1).  If the neutral fermion $\chi$ is dark matter, then whether 
it is Majorana or Dirac will have very different predictions, especially 
in the dark-matter bremsstrahlung process of $\chi \chi \to e^+ e^- \gamma$. 
As Fig.~5 shows, Dirac dark matter seems to be favored by current data.

\underline{\it Acknowledgements}~:~ Q.H.C.~is supported in part by the 
Argonne National Laboratory and University of Chicago Joint Theory Institute 
(JTI) Grant 03921-07-137, and by the U.S.~Department of Energy under 
Grants No.~DE-AC02-06CH11357 and No.~DE-FG02-90ER40560.  
E.M.~is supported in part by the U.~S.~Department of Energy under Grant 
No.~DE-FG03-94ER40837.  
G.S. is supported in part by the U.~S.~Department of Energy under Grant 
No.~DE-AC02-06CH11357.


\end{document}